\documentclass{acm_proc_article-sp}
\newcommand{\OoT}{Out-of-Turn}
\newcommand{\Oot}{Out-of-turn}
\newcommand{\oot}{out-of-turn}

\begin{document}

\title{Automatically Generating Interfaces for\\Personalized Interaction with Digital Libraries}

\numberofauthors{3}

\author{Saverio Perugini, Naren Ramakrishnan, and Edward A. Fox\\
Department of Computer Science\\
Virginia Tech, Blacksburg, VA 24061, USA\\
E-mail: \{sperugin, naren, fox\}@cs.vt.edu}

\date{February 10, 2004}
\maketitle

\begin{abstract}
We present an approach to automatically generate interfaces supporting
personalized interaction with digital libraries; these interfaces
augment the user-DL dialog by empowering the user to (optionally) supply
out-of-turn information during an interaction, flatten or restructure
the dialog, and enquire about dialog options. Interfaces generated
using this approach for CITIDEL are described.
\end{abstract}

\vspace{-0.10cm}



\section{Introduction}
There have been many tools proposed recently
to assist librarians, who might have little or no experience in building
computer systems, in constructing digital libraries (e.g., 
Greenstone~\cite{Greenstone} and 5SGraph~\cite{5SGraph}).
We present here a system for automatic generation of {\it interfaces} 
to digital libraries, especially those that support personalized
interaction. We model our system after the `staging transformations'
framework~\cite{imWWW2004} that supports functional specification and realization
of user-DL dialogs.

Our view of personalized DL interaction is where the user and DL
take turns exchanging initiative and where the user is empowered to
supply partial information in as expressive a manner as possible.
Today's DL interfaces are predominantly hyperlink driven; to reconcile
the mismatch between the hardwired hyperlink structure  and user's
information seeking goals, we provide the capability to have an
{\it out-of-turn interaction}, where the user can supply some
unsolicited, but relevant, information to the DL. When such an
\oot\ input is made, the user momentarily takes the initiative
and the site restructures itself before reclaiming the initiative.
Since \oot\ interaction is optional and unintrusive, it can
be introduced at multiple times, at the user's discretion.
\Oot\ interaction is targeted at focused dialogs where the user
has a specific information-seeking goal in mind but the DL's
current hyperlink structure does not accommodate it (e.g., a
user interested in papers by `Belkin' but is unsure what categories
Belkin has published in).

We have built two interfaces to support \oot\ interaction in DLs:
a toolbar embedded into a web browser for \oot\ textual input,
and voice-enabled content pages for \oot\ speech input.
Studies using these techniques have revealed that users are adept
at recognizing when \oot\ interaction is necessary but have also
highlighted the need for supplemental operators to enhance the personalized
experience~\cite{ExtemporeTR}. 
While some of the operators are
always applicable, others are only defined under certain conditions (see below),
hence the need to develop
customized interaction interfaces for specific DLs.  

\section{Generating Personalized\\DL Interfaces}
Our generator presumes a DL modeling such as in the OAI-augmented
5S framework or an XML-based representation. We currently target 
DL pages indexed according to some classification scheme inherent
in the representation; the classification
terms are available for \oot\ input as well as additional terms
modeled in the leaf documents (and not explicitly used in the
classification). The generator currently supports the
following personalized interactions:

\vspace{-0.50cm}

\begin{itemize}
\item {\bf Basic out-of-turn interaction:} This interaction
technique prunes all paths through the classification scheme
that do not contain the out-of-turn input. A combination of
program slicing transformations is utilized to identify the
paths which must be retained. This technique only allows
partial input included in the classification scheme.

\item {\bf Generalized out-of-turn interaction:} If the user's
input does not correspond to a label in the classification scheme,
this interaction technique is more appropriate as it can accommodate
partial information modeled at the leaves, in addition to
the tree labels.

\item {\bf Meta-enquery: what may I say?:}
Keeping users abreast of what information is
available to communicate is a feature all DLs must address.  This 
technique permits the user to enquire about the partial information
that remains unspecified in the user-DL dialog.

\item {\bf Collect results:} This is a dialog termination technique
and allows the user to request that the classification be flattened
and re-presented as a flat list of relevant pages.

\item {\bf Restructure classification:} For DLs with categorical
facets, this interaction technique enables the creation of a personalized
browsing hierarchy, which can be further navigated in a dialog, e.g., a
DL organized along a author-journal-title motif could be restructured into
a journal-author-title motif, supporting interactive aggregation scenarios.

\end{itemize}

\vspace{-0.4cm}

\section{Example}

\begin{figure}
\centering
\includegraphics[width=8.4cm]{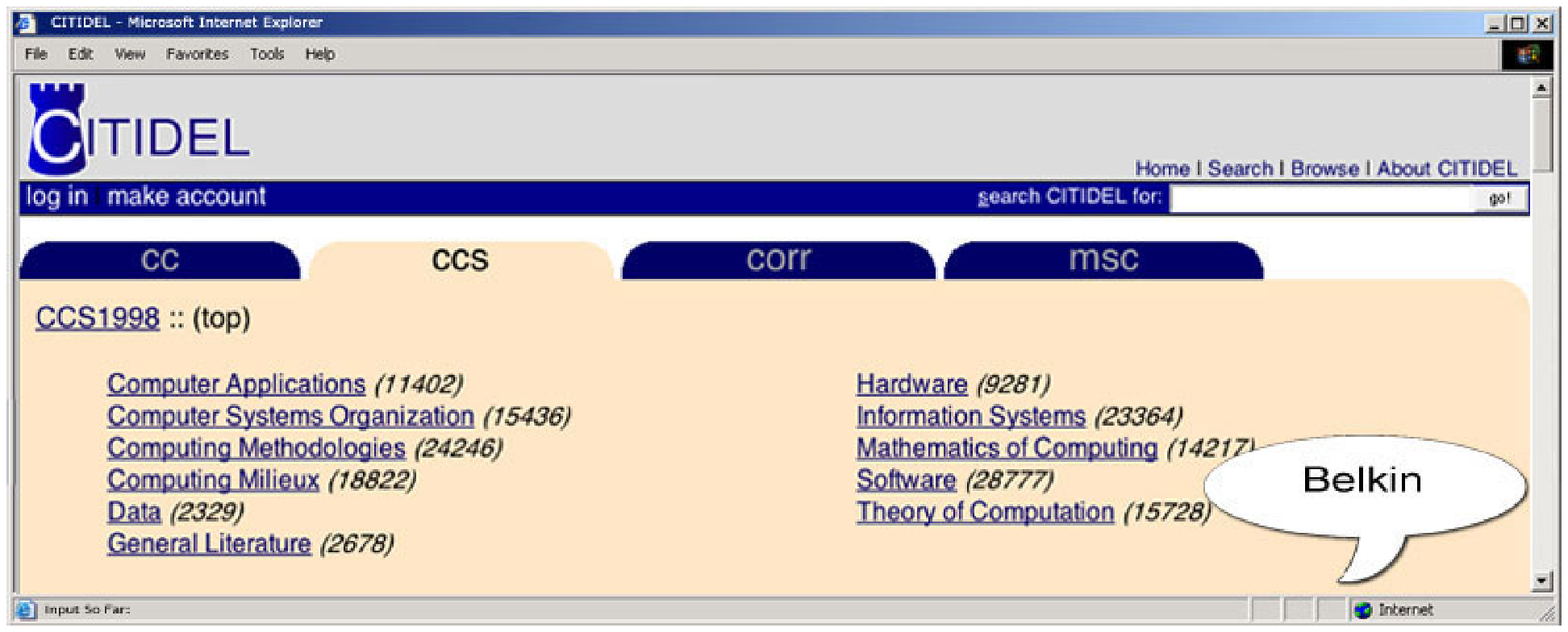}\\
$\Downarrow$\\
\includegraphics[width=8.4cm]{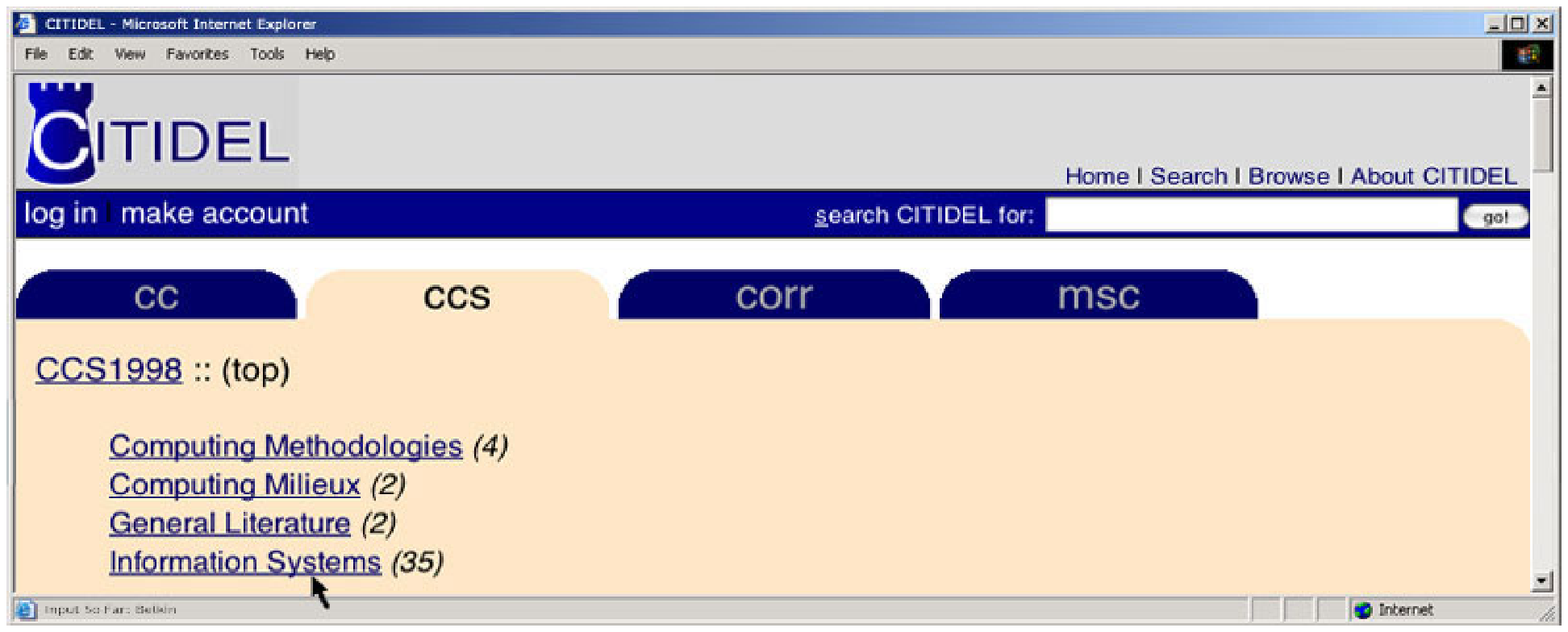}\\
$\Downarrow$\\
\includegraphics[width=8.4cm]{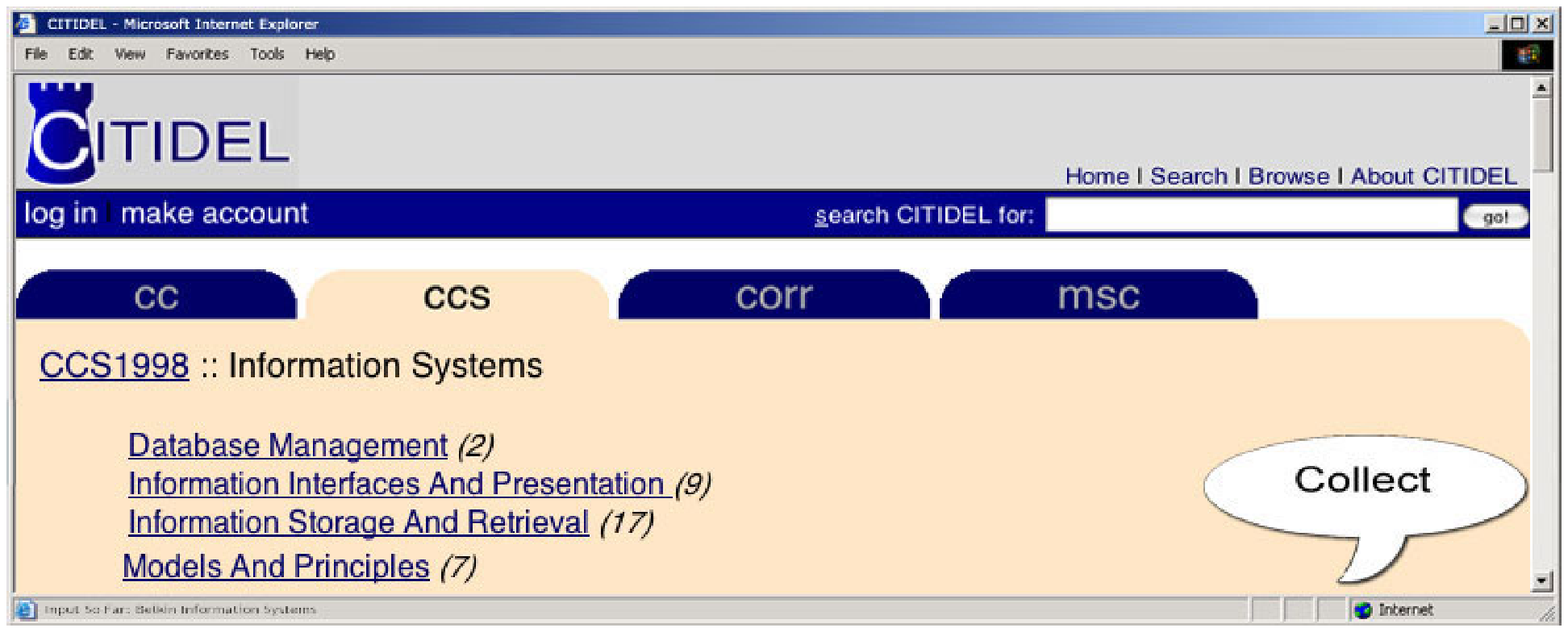}\\
$\Downarrow$\\
\includegraphics[width=8.4cm]{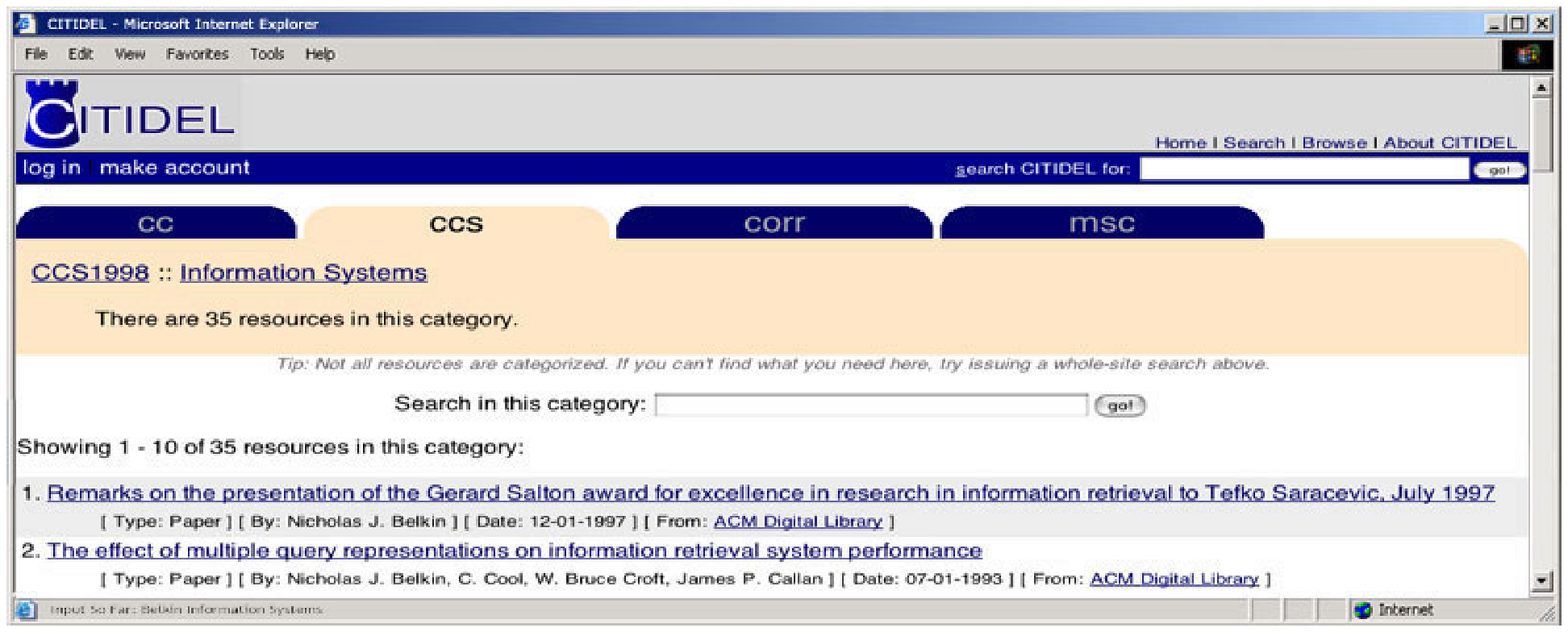}\\
\caption{An example of a personalized interaction with CITIDEL.}
\label{sequence}
\end{figure}

Fig.~\ref{sequence} illustrates a personalized DL interaction with
CITIDEL (Computing and Information Technology Interactive Digital Educational 
Library; citidel.org)~\cite{CITIDEL-JCDL2004TR}
using \oot\ speech input; the user is initially unable to respond to the
solicitation of literature category and instead says `Belkin' out-of-turn~(screen i).
This causes many leaf pages to be removed~(notice reduced frequency purviews annotating each
hyperlink label) and
some categories~(e.g.,~`Hardware') to be completely pruned out, since
Belkin's papers are not indexed under such categories~(screen ii). The user then
responds to the initiative by following the `Information Systems' hyperlink~(screen iii).
After this point, the user decides to terminate the dialog and request a flat
list of Belkin's papers~(screen iv).

Our approach to automatic interface generation produces either a 
browser toolbar in XUL
(XML User interface Language) for use in the cross-platform Mozilla web browser
or a SALT (Speech Language Application Tags) voice interface for use in
Internet Explorer. 
Currently, the toolbar captures \oot\ input as a bag of words (hence supporting
phrases) whereas the voice-enabled pages do not accommodate lengthy
constructs (for ease of speech recognition).

The generator is built in Java; a graphical interface (not shown here) 
supports configuration of DL interfaces by allowing designers to choose
the interaction techniques they desire to support.
For librarians who are more computer-literate, we defined a small markup
language~(OTML--\OoT\ Markup Language) using XSchema and
employed the translation capabilities of XSLT to compile it into a XUL toolbar or
SALT voice interface.  Use of OTML supports a level of customization finer than
that of the generator (e.g., designers can customize tooltips for each widget in the
generated toolbar).

CITIDEL interfaces generated by our tool enable multimodal
personalized interaction and have had qualified
success~\cite{CITIDEL-JCDL2004TR}.  We expect approaches to generating
interfaces for DLs such as ours to become more popular with the rise in
end-user programming.

\bibliographystyle{abbrv}
\bibliography{jcdl2004}

\balancecolumns

\end{document}